\documentstyle {article}

\renewcommand{\H}{{\cal H}}

\newcommand{\M}{{\cal M}}
\newcommand{\K}{{\cal K}}
\newcommand{\B}{{\cal B}}

\newcommand{\A}{{\cal A}}

\newcommand{\U}{{\cal U}}

\newcommand{\C}{{\bf C}}
\newcommand{\R}{{\bf R}}

\begin {document}
\newtheorem{definition}{Definition}
\newtheorem{remark}{Remark}
\sloppy
\large

\begin {center}
{\bf KMS STATES ON THE SQUARE OF WHITE NOISE ALGEBRA}
\end {center}

\bigskip

\begin {center}
{\bf Luigi Accardi, Grigori Amosov, Uwe Franz}
\end {center}

\medskip

{\bf Abstract.} It was shown in $[AFS00]$ that there are only three
types of irreducible unitary representations $\theta $ of
$\bf sl_2$. Using the Schurmann triple one can associate with
each $\theta $ a number of representations of the square of
white noise (SWN) algebra $\A$.
However, in analogy with the Boson, Fermion and
q-deformed case, we expect that some interesting non irreducible
representations of $\bf sl_2$ may result in
GNS representations of
KMS states associated with some evolutions on $\A$. In the present
paper determine the structure of the $*$--endomorphisms of the SWN
algebra, induced by linear maps in the 1--particle Hilbert algebra,
we introduce the SWN analogue of the quasifree evolutions and find
the explicit form of the KMS states associated with some of them.
\medskip

{\bf 1. The square of white noise and its representations.}

\begin{definition}\label{swual} The square of white noise (SWN) algebra $\A$
over the Hilbert algebra $\K=L^2(\R)\cap L^{\infty }(\R)$ (see [ALV99,
AFS00]),
is the unital $*$--Lie algebra with generators 1 (central element),
$b_{\phi },b_{\phi }^+,n_{\phi},\ \phi \in \K$, which are linearly
independent (in the sense that $f_01+b_{f_1}+b_{f_2}+n_{f_3}=0$ with
$f_0\in{\bf C}$ and $f_j\in{\cal K}$ $(j=1,2,3)$ if and only if
$f_1=f_1=f_2=f_3=0$) and relations
$$
[b_{\phi },b_{\psi }^+]=\gamma <\phi ,\psi >1+n_{\overline {\phi }\psi },
\eqno (SWN_1)$$
$$[n_{\phi },b_{\psi }]= -2b_{\overline {\phi }\psi },
\eqno (SWN_2)$$
$$[n_{\phi },b_{\psi }^+]=2b_{\phi \psi }^+,
\eqno (SWN_3)$$
$$(b_{\phi })^*=b_{\phi }^+,\ (n_{\phi })^*=n_{\overline {\phi }},
\eqno (SWN_4)$$
$$[n_\phi,n_\psi]=[b^+_\phi,b^+_\psi]=[b_\phi,b_\psi]=0\eqno(SWN_5)$$
where $\gamma $ is a fixed strictly positive real parameter (coming from
the renormalization), $<\cdot ,\cdot >$ is a inner product in $\K $
and $\phi ,\psi \in \K$. Furthermore $b^+$ and $n$ are linear and $b$
is anti-linear in the functions from $\K$ and we assume that
$$
X_\phi=0\quad\hbox{if and only if }\psi=0
\eqno (SWN_6)
$$
\end{definition}

\begin {remark}
The involution on ${\cal K}$ will be
indifferently denoted $f\mapsto f^*$ or $f\mapsto\overline f$.
Following [AFS00] all the irreducible unitary representations
of $\A$ can be obtained as follows.
Consider the Lie algebra $\bf sl_2$ with generators $\{B^+,B^-,M\}$,
relations
$$[B^-,B^+]=M,\ [M,B^{\pm}]=\pm B^{\pm},$$
and involution $(B^-)^*=B^+,\ M^*=M$.
Let $\rho$ be an irreducible representation of $\bf sl_2$
and
therefore of its universally enveloping unital algebra
$\U({\bf sl_2})$ in a Hilbert space
$\H_0$ and $\eta$ be a 1--cocycle for $\rho$.
Define $L(uv)=<\eta (u^*),\eta (v)>,\ u,v\in \U_0({\bf sl_2})$,
where $\U_0({\bf sl_2})$ is the algebra $\U({\bf sl_2})$ without unit.
Then $(\rho ,\eta ,L)$ is called a Sch\"urmann triple and
a representation $\pi $ of $\A$ in the Hilbert space
$\Gamma (\H_0\otimes L^2(\R))$ can be obtained from the relations
$$
\pi (b_{\chi _{[s,t[}})=\Lambda _{st}(\rho (B^-))+A^*_{st}(\eta (B^-))+
A_{st}(\eta (B^+))+L(B^-)(t-s)Id,
$$
$$
\pi (b_{\chi _{[s,t[}}^+)=
\Lambda _{st}(\rho (B^+))+A^*_{st}(\eta (B^+))+
A_{st}(\eta (B^-))+L(B^+)(t-s)Id,
$$
$$
\pi (n_{\chi _{[s,t[}})=
\Lambda _{st}(\rho (M))+A^*_{st}(\eta (M))+
A_{st}(\eta (M))+(L(M)-\gamma )(t-s)Id,
\eqno (Rep)
$$
where $\Lambda _{st}=\Lambda _t-\Lambda_s,
A_{st}^*=A_t^*-A_s,A_{st}=A_t-A_s$ are the
conservation, creation and annihilation processes on the symmetric
Fock space $\Gamma (\H_0\otimes L^2(\R))$ satisfying the relations
(see [Par92, Mey95])
$$[A_t(\phi),A_s^*(\psi)]=(t\wedge s)<\phi ,\psi>,$$
$$[A_t(\phi ),A_s(\psi )]=[A_t^*(\phi ),A_s^*(\psi )]=0,$$
$$[\Lambda _t(X),\Lambda _s(Y)]=\Lambda _{t\wedge s}([X,Y]),$$
$$[\Lambda _t(X),A_s(\phi )]=-A_{t\wedge s}(X^*\phi ),
[\Lambda _t(X),A_s^*(\phi )]=A^*_{t\wedge s}(X\phi ),
\eqno (CCR)$$
$$\phi ,\psi \in \K ,X,Y\in {\cal B}(\K),$$
$\chi _I\ (I\subset \R)$ denotes the multiplication operator by
the characteristic function of $I$ ($=1$ on $I$ and $=0$ outside $I$),
$\gamma $ is the same as in $(SWN_1)$.
Conversely, all the irreducible representations of $\A$ arise in this way.
We shall say that the representation
$\rho $ is associated with the representation $\pi$ anche the cocycle
$\eta$. Hence to classify the representations of $\A$ one needs to
investigate
the representations of $\U(\bf sl_2)$ and their cocycles.
\end {remark}

The contents of the present paper is the following.
In Section 2 we introduce the quasifree evolutions on $\A$.
Then we obtain a classification of these evolutions and prove
their spatiality. In Section 3 we consider the KMS states
associated with a special subclass of the quasifree evolutions
on $\A$.

\medskip

{\bf 2. Endomorphisms of the SWN algebra.}

\medskip
\begin{definition}\label{hilbalg} A Hilbert algebra is a *--algebra
${\cal K}$, not necesarily with unit, endowed with a
scalar product $\langle f,g\rangle\in{\bf C}$ satisfying $\langle
f,gh\rangle=\langle g^*f,h\rangle$ $(f,g,h\in{\cal K})$.
A Hilbert algebra endomorphism (resp. automorphism) of ${\cal K}$ is
a *--endomorphism
(resp. *--automorphism) $T$ of the *--algebra structure
$$ T(\phi \psi )=T(\phi )T(\psi ),\ (T(\phi ))^*=T(\overline {\phi}) $$
which is also an isometry (resp. unitary operator), of the pre--Hilbert
space structure in the sense that, $\forall\,\phi$, $\psi\in{\cal K}$
$$
\langle T(\phi),T(\psi )\rangle =\langle \phi , \psi\rangle
\eqno (1)
$$
\end{definition}

\noindent{\bf Theorem 1.} {\it
Let $T^1,T^2,T^3$ be linear operators on $\K $. Define a map
$\tau '$ acting on the generators $b,b^+,n$ by the formula
$$
1\to1;\ b_{\phi }\to b_{T^1\phi },\ b^+_{\phi }\to b^+_{T^2\phi },\
n_{\phi }\to n_{T^3\phi }.
\eqno (2)
$$
The map $\tau'$ can be extended to a
$*$--endomorphism of $\A$ if and only if there exist a Hilbert algebra
endomorphism $T$ of $\K$ and a real-valued function $\alpha $ on $\R,$ such
that, for any $\phi\in \K$,
$$
T^3(\phi )=T(\phi )
\eqno (3)
$$
$$
T^1(\phi )=T^2(\phi )=e^{i\alpha }T(\phi )
\eqno (4)
$$
The endomorphism $\tau'$ is a an
automorphism if and only if $T$ is an automorphism.
}

{\bf Proof}. Because of the linear independence of the generators and of
$(SWN_6)$, it follows from the relation $(SWN_1)$ that
$$(T^1)^*T^2=Id,\ T^3(\overline {\phi }\psi )=\overline
{T^1(\phi )}T^2(\psi ),
\eqno (5)$$
where the first identity in condition (5) has to be interpreted in the
sense of (1).
From (SWN$_2$) and (SWN$_6$) it follows that
$$T^1(\overline {\phi }\psi )=\overline {T^3(\phi )}T^1(\psi ),\eqno
(6)$$ from $(SWN_3)$ and (SWN$_6$), that $$T^2(\phi \psi
)=T^3(\phi )T^2(\psi )\eqno (7)$$ and from (SWN$_4$) and (SWN$_6$),
that $$T_0:=T^1=T^2,\ \overline {T^3(\phi )}=T^3(\overline {\phi
}),\eqno (8)
$$ for any $\phi ,\psi \in \K$.

Therefore (5),(6,(7),(8) are respectively equivalent to:
$$
T^*_0T_0=id\eqno (9)
$$
$$
T^3(\overline\phi)=\overline{T^3(\phi)}\eqno (10)
$$
$$
T^3(\overline\phi\psi)=\overline{T_0(\phi)}T_0(\psi)\eqno (11)
$$
$$
T_0(\phi\psi)=T^3(\phi)T_0(\psi)\eqno (12)
$$

From (11) with $\phi=\psi$, we deduce that for any
$\psi\in{\cal K}$
$$
T^3(|\psi|^2)=|T_0(\psi)|^2\eqno (13)
$$
Again from (11), with $\psi$ replaced by $\psi\chi$ we obtain
$$T^3(\overline\phi\psi\chi)=\overline{T_0(\phi)}T_0(\psi\chi)$$
and, from (12) this is equal to
$$\overline{T_0(\phi)}T^3(\psi)T_0(\chi)$$
Choosing $\phi=\chi$ and using (13) we deduce
$$T^3(|\phi|^2\psi)=|T_0(\phi)|^2T^3(\psi)=T^3(|\phi|^2)T^3(\psi)$$

Since any positive element in ${\cal K}$ can be written in the form
$|\phi|^2$ for some $\phi\in{\cal K}$, this implies that, for any
positive element $\phi$ in ${\cal K}$
$$
T^3(\phi\psi)=T^3(\phi)T^3(\psi)\eqno (14)
$$
Since any $\phi\in{\cal K}$ is linear combination of positive elements
we conclude that (14) holds for any $\phi$, $\psi\in{\cal K}$.

Combining (13) and (14) we conclude that, for any
$\psi\in{\cal K}$
$$|T^3(\psi)|^2=|T_0(\psi)|^2$$
Thus for each $\psi\in{\cal K}$ there exists a real valued measurable
function $\alpha_\psi$ such that
$$
T_0(\psi)=e^{i\alpha_\psi}T^3(\psi)\eqno (15)
$$

Finally (15) and (9) imply that, for any $\phi\in{\cal
K}$
$$\langle T^3(\phi),T^3(\phi)\rangle=\langle T_0(\phi),T_0(\phi)
\rangle=\langle\phi,\phi\rangle$$
and, by polarization, this implies that $T^3$ is isometric. Thus
$T^3$ is a Hilbert algebra endomorphism. Let us denote
$$T:=T^3$$
If for some $t>0$, supp$\,(\varphi)\subseteq[-t,t]$ then (15)
implies that
$$e^{\alpha_\varphi} T(\varphi)=T_0(\varphi)=T_0(\varphi\chi_{[-t,t]})=
T(\varphi)T_0(\phi_{[-t,t]})=
$$
$$
e^{i\alpha\chi_{[-t,t]}T(\varphi)T
(\chi_{[-t,t]})}=e^{i\alpha\chi_{[-t,z]}}T(\varphi)$$
Thus, for any $\varphi\in{\cal K}$ with
supp$\,(\varphi)\subseteq[-t,t]$, one has, on supp$\,(T(\varphi))$:
$$
\alpha_\varphi=\alpha_{\chi_{[-t,t]}}\quad;\qquad\hbox{a.e.}\eqno
(16) $$

In particular, if $s\leq t$
$$
\alpha_{\chi_{[-s,s]}}=\alpha_{\chi_{[-t,]}}\ ;\quad\hbox{a.e. on supp }
T(\chi_{[-s,s]})\eqno (17)
$$
Since $T$ is an endomorphism of $K$, $T(\chi_{[-t,t]})$ is a
self--adjoint projection in ${\cal K}$, hence it has the form
$$T(\chi_{[-t,t]})=\chi_{I_t}$$
for some measurable subset $I_t\subseteq{\bf R}$.
By the isometry property one can suppose that $t\mapsto I_t$ is
increasing:
$$s\leq t\Rightarrow I_s\subseteq I_t\ ;\quad\hbox{a.e.}$$
Denote
$$\chi_I:=\sup\chi_{I_t}=\chi_{\sqcup_{t\geq0}I_t}$$
since the union can be taken on any sequence increasing to $+\infty$,
one can assume that $I$ is measurable up to a set of measure zero.
Moreover for any sequence $t_n\uparrow+\infty$ the function
$$\alpha:=\lim\alpha_{\chi_{[-t_n,t_n]}}$$
is well defined on $I$ and measurable up to a sub--set of measure zero
of $I$. Because of (16) for any function $\varphi$ with bounded
support
one has
$$T_0(\varphi)=e^{i\alpha_\varphi}T(\varphi)=e^{i\alpha}T(\varphi)$$
Since $T_0$ is isometric and the functions with bounded support are
dense in ${\cal K}$, it follows that
$$
T_0=e^{i\alpha}T\eqno (18)
$$
Conversely, if $T:{\cal K}\to{\cal K}$ is a Hilbert algebra
endormorphism, $\alpha:{\bf R}\to{\bf R}$ a measurable function and $T_0$,
$T_1$, $T_2$ are defined by (18), (8) respectively, then
the map $\tau'$, defined by (2) preserves the commutation
relations of the SWN, hence is a $*$--Lie algebra endomorphism.

Finally it is clear that $\tau'$ will be an automorphism (i.e.
surjective endomorphism) if and only if $T$ is on to, i.e. unitary.

\noindent{\bf Example}. A non surjective endomorphism of ${\cal A}$

Consider the following isometry on $L^2({\bf R})$:
$$
Vf(x)=\cases{
0\quad\hbox{if}\quad x\in[0,1)\cr
f(x-1)\quad\hbox{if}\quad x\in[1,+\infty)\cr
0\quad\hbox{if}\quad x\in(-1,0]\cr
f(x+1)\quad\hbox{if}\quad x\in(-\infty,-1]\cr}\eqno (19)
$$
One has:
$$\langle Vf,Vg\rangle=\int[Vf(x)]^-[Vg(x)]dx$$
$$=\int^\infty_1\overline f(x-1)g(x-1)dx+\int^{-1}_{-\infty}\overline
f(x+1)g(x+1)dx=$$
$$=\int^\infty_0\overline f(y)g(y)dy+\int^0_{-\infty}\overline f(y)g(y)
dy=\int^{+\infty}_{-\infty}\overline f(y)g(y)dy=\langle
f,g\rangle_{L^2}$$
Thus $V$ is isometric:
$$V^*V=1$$
However
$$\hbox{Range }V=\{f\in L^2({\bf R}):f(x)=0\hbox{ if }x\in[-1,1]\}$$
which is a proper subspace of $L^2({\bf R})$.

\begin{remark} Notice that $V$ maps
$L^2\cap L^\infty({\bf R})=:{\cal K}$
into itself and clearly induces a Hilbert algebra endomorphism of ${\cal
K}$. Let us denote it $T^3$. Notice however that, if $\chi$ is any
characteristic function of a bounded subset in ${\bf R}$, then
$$T^3(\chi)\leq\chi_{(-1,1)^c}$$
Therefore
$$\overline\chi=\sup\{T^{(3)}\chi:\chi\hbox{ projectors on bounded
subsets of } {\bf R}=\chi_{(-1,1)^c}<1$$
\end{remark}

\noindent{\bf Definition 3.} {\it The endomorphisms of $\A$,
introduced in Theorem 1 are called quasifree. If $\tau '$ is such an
endomorphism and $(T,\alpha )$ is the pair associated to it via
Theorem 1, we shall say that $\tau '$ is obtained by the lifting (or
second quantization) of the pair $(T,\alpha )$.}

The theorem implies that every quasifree endomorphism $\tau '$ of $\A$
can be represented as a composition $\tau '=\tau ^1\tau ^2$ such that

$\bf (A)$ $\tau ^1$ is obtained by the lifting of the pair $(T,0)$;

$\bf (B)$ $\tau ^2$ is obtained by the lifting of the pair $(1,\alpha)$.

A quasifree endomorphism will be called of type $(A)$, resp.
$(B)$ if it belongs to one of these classes.
Notice that, if we consider a one-parameter group of automorphisms with
parameter set $\R$, then the associated family of operators
$(T_t)_{t\in \R}$ must be a group of
endomorphisms in case of type (A) and the function $\alpha _t(x)$
must be of the form
$\alpha _t(x)=\alpha (x)t,\ t\in \R$,
in case of type (B).
We are interested to study the spatiality of these evolutions relatively
to
the representations of $\A$.\medskip

\noindent {\bf Proposition 1.} {\it Let $\pi $ be a representation of $\A$
in the Hilbert space
$\H=\Gamma (\H_0\otimes L^2(\R))$ generated by the Schurmann
triple $(\rho ,\eta ,L)$. Given
a type $\bf (A)$ quasifree automorphism $\tau ' $
of the SWN algebra $\A$,
there is an automorphism $\tau $
of the algebra $\B(\H)$ such that
$\tau (\pi (x))=\pi (\tau '(x)),\ x\in \A$.}\medskip

\noindent{\bf Proof}. Let the automorphism $\tau '=\tau '_T$ be obtained by
the lifting of the pair $(T,0)$.
To prove the proposition we must look for a unitary operator $\U$
in $\H$ implementing $\tau'$. Every automorphism of $\K$ can
be continued to an automorphism of the von Neumann algebra
$L^{\infty }(\R)$ that is unitary implementable. Therefore
there is a unitary operator $U$ acting in $L^2(\R)$ such that
$T(x)=UxU^*,\ x\in \K$. The formula $\U e(f\otimes \phi )=
e(f\otimes U\phi ),\ f\in \H_0,\ \phi \in \Gamma (L^2(\R))$,
defines a unitary operator $\U=\Gamma(1\otimes U)$ on the exponential
vectors
$e(f\otimes \phi )$ in $\H$ and therefore on the whole of ${\cal H}$.
The property $\pi (\tau '(x))=\U \pi (x)\U ^*,\ x\in \A,$ holds.
One can define
$\tau (x)=\tau _T (x) = \U x\U ^*,\ x\in \B(\H)$.
$\Box$

Given a type $\bf (B)$ quasifree automorphism $\tau ' $
of the SWN algebra $\A$ obtained by the lifting of
pair $(1,\alpha )$ and a representation $\pi $
of $\A$ generated by the Schurmann triple
$(\rho ,\eta ,L)$, it is natural to conjecture that the associated
automorphism $\tau (\cdot)$, of the algebra $\B(\H)$,
$(\tau (\pi (x))=\pi (\tau '(x)),\ x\in \A)$, has the form
$\tau(\cdot)=e^{iH}(\cdot)e^{-iH}$ with
$$
H=\frac {1}{2}\int \alpha (t)d n_t,
$$
$$
dn_t=d\Lambda (\rho (M))+dA^*(\eta
(M))+dA
(\eta (M))+(L(M)-\gamma)dt\eqno (20)
$$
This conjecture can be proved by an approximation argument which makes
use of the following.\medskip

\noindent{\bf Proposition 2}. {\it In the above notations let $\alpha$ be a
locally constant function vanishing outside a bounded interval and let
$(\tau_\varepsilon)$ be the 1--parameter automorphism group of ${\cal
A}$ associated to the pairs $(1,\varepsilon\alpha),\
\varepsilon\in{\bf R}$. Then for any $x\in{\cal A}$
$$
\tau_\varepsilon(\pi(x))=e^{i\varepsilon H}\pi(x)e^{-i\varepsilon H}
\eqno (21)
$$
}
\medskip

\noindent{\bf Proof}. For each $x\in{\cal A}$ denote
$$x(\varepsilon)=\tau_\varepsilon(x)$$
Choosing $x=b^+_\psi$ $(\psi\in{\cal K})$ one has
$$\partial_\varepsilon\pi(b^+_\psi(\varepsilon))=\partial_\varepsilon\pi
(b^+_{e^{i\varepsilon\alpha}\psi})=\pi(b^+_{i\alpha e^{i\varepsilon
\alpha}\psi})=i\alpha\pi(b^+_\psi(\varepsilon))$$
where in the last identity we have used the fact that, because of the
independent increment property, it is sufficient to consider the case
$\alpha=$ constant on the support of $\psi$.

In our assumptions the operator $H=H(\alpha)$ is self--adjoint and
$$C^+_\psi(\varepsilon):=e^{i\varepsilon H_\pi}(b^+_\psi)e^{-i\varepsilon
H}$$
satisfies the equation
$$\partial_\varepsilon C^+_\psi(\varepsilon)=ie^{i\varepsilon H}
[H,\pi(b^+_\psi)]e^{-i\varepsilon H}=ie^{i\varepsilon H}\pi
([n_{\alpha/2},b^+_\psi])e^{-i\varepsilon H}$$

It follows that $\tau_\varepsilon(b^+_\psi))$ and
$C^+_\psi(\varepsilon)$ satisfy the same ordinary differential equation
with the same initial condition. Since $\alpha$ is constant, in both
cases the unique solution is
$$\tau_\varepsilon(\pi(b^+_\psi))=C^+_\psi(\varepsilon)=
\pi(b^+_{e^{i\varepsilon \alpha}\psi})$$
Similarly one verifies that
$$\tau_\varepsilon(\pi(b_\psi))=\pi(b_{e^{i\varepsilon\alpha}\psi})\ ;
\quad\tau_\varepsilon(\pi(n_\psi))=\pi(n_\psi)$$
and from this (21) follows.\bigskip\medskip

\noindent{\bf 3. KMS states associated with quasifree evolutions on
the SWN algebra}

\medskip
Let $\tau '_{\lambda }=(\tau _t')_{t\in \R}$ be a group of
type (B) quasifree automorphisms on $\A$ defined by
$$
\tau _t'(b_{\phi })=\lambda ^{-it}b_{\phi },\
\tau _t'(b^+_{\phi })=\lambda ^{it}b^+_{\phi },
$$
$$
\tau _t'(n_{\phi })=n_{\phi },\
t\in \R,\ 0<\lambda <1.
$$
Pick up the representations $\rho ^{\pm}$ of $\U({\bf sl_2})$
in $l^2$ introduced in $[AFS00]$ as
$$
\rho ^+(B^+)e_n=\rho ^-(B^-)e_n=\sqrt {(n+1)(n+2)}e_{n+1},
$$
$$
\rho ^+(B^-)e_n=\rho ^-(B^+)e_n=\sqrt {n(n+1)}e_{n-1}
$$
$$
\rho ^{\pm}(M)e_n=\pm (2n+2)e_n,
$$
where $\{e_0,e_1,\dots ,e_n,\dots \}$ is an orthonormal basis of
$l^2$. Then define a pair of states $\phi ^{\pm }=\phi ^{\pm}
_{\lambda }$ on $\U({\bf sl_2})$ by the formula
$$
\phi ^{\pm }_{\lambda
}(\cdot )=(1-\lambda ) \sum \limits _{n=0}^{+\infty } \lambda ^n (e_n
,\rho ^{\pm}(\cdot )e_n).
$$
In the representation $\pi =\rho ^+\otimes \rho ^-$ of $\U({\bf sl_2})
\otimes \U({\bf sl_2})$
in the Hilbert space $\H_{\phi }=l^2\otimes l^2$ we get
$$\phi^+_\lambda(x)=(\psi _{\lambda },\pi (x\otimes {\bf 1})\psi
_{\lambda }),$$ $$\phi^-_\lambda(x)=(\psi _{\lambda },\pi ({\bf
1}\otimes x)\psi _{\lambda }),\ \phi_\lambda (x\otimes
y)=(\psi_\lambda,\pi(x\otimes y)\psi_\lambda)$$ $$x,y\in \U({\bf
sl_2}),$$ where $\psi_\lambda=\sqrt {1-\lambda }\sum
\limits _{n=0}^{+\infty } \lambda ^{\frac {n}{2}}e_n\otimes e_n\in
\H_{\phi}$. So one can consider $\pi $ as the GNS representation
associated with the state $\phi_\lambda$.  Put $B_1^+=\pi
(B^+\otimes {\bf 1}),\ B_1^-=\pi (B^-\otimes {\bf 1}),\ M_1=\pi
(M\otimes {\bf 1})$ and $B_2^-=\pi ({\bf 1}\otimes B^-),\ B_2^+=\pi
({\bf 1}\otimes B^+),\ M_2=\pi ({\bf 1}\otimes M)$. Every triple
$(B_i^+,B_i^-,M_i)$ satisfies the relations of $\bf sl_2$
by the definition.\medskip

{\bf Proposition 3.} {\it The pairwise
commuting operators $(B_1^+,B_1^-,M_1)$
and $(B_2^+,B_2^-,M_2)$ satisfy the following relations
$$
B_1^-\psi _{\lambda }=\sqrt {\lambda }B^-_2\psi _{\lambda },\
B_1^+\psi _{\lambda }= \frac {1}{\sqrt {\lambda }}B_2^+\psi
_{\lambda },
$$
$$ M_1\psi _{\lambda }=-M_2\psi _{\lambda }.  $$ }\medskip

\noindent{\bf Proof}. The identities
$$
B^+_1e_n\otimes e_n=
\pi (B^+\otimes {\bf 1})e_n\otimes e_n=
$$
$$
=\rho ^+(B^+)e_n\otimes e_n=\sqrt {(n+1)(n+2)}e_{n+1}\otimes e_n=
$$
$$
=e_{n+1}\otimes \rho ^-(B^+)e_{n+1}=\pi ({\bf 1}\otimes B^+)e_{n+1}
\otimes e_{n+1}=B^+_2e_{n+1}\otimes e_{n+1}
$$
imply
$$
B^+_1\psi _{\lambda }=\sqrt {1-\lambda }\sum \limits
_{n=0}^{+\infty } \lambda ^{\frac {n}{2}}B^{+}_1e_n\otimes e_n=
\sqrt {1-\lambda }\sum \limits _{n=1}^{+\infty } \lambda ^{\frac
{n-1}{2}}B^{+}_2e_n\otimes e_n= $$ $$ =\frac {1}{\sqrt {\lambda
}}B^+_2\psi _{\lambda }, $$ where we used the formula
$B^+_2e_0\otimes e_0=e_0\otimes \rho ^-(B^+)e_0=0$. The remaining
equalities can be proved in the same way. $\Box$

Consider the Schurmann triple $(\pi
,\eta_\lambda,\tilde \phi_\lambda)$ consisting of the representation
$\pi $ of $\U({\bf sl_2})\otimes \U({\bf sl_2})$, the trivial cocycle
$\eta_\lambda(x)=\pi (x)\psi_\lambda$ and the conditionally positive
functional
$\tilde \phi _{\lambda }(x)= (\psi ,(\pi (x)-\varepsilon (x))\psi
),\ x\in \U({\bf sl_2})\otimes \U ({\bf sl_2})$, where $\varepsilon $
is a counit. The restrictions of
$(\pi ,\eta_\lambda, \tilde \phi_\lambda)$ give us two Schurmann
triples $(\pi _{\pm},\eta _{\lambda,\pm}, \tilde \phi
_{\lambda,\pm})$ consisting of representations $\pi _{\pm}$ of
$\U(\bf sl_2)$ by $(B_1^+,B_1^-,M_1)$ and $(B_2^+,B_2^-,M_2)$
correspondingly such that $\pi _+(x)=\pi (x\otimes {\bf 1}),\ \pi
_-(x)= \pi ({\bf 1}\otimes x)$, $\eta _{\lambda,\pm}(x)=\pi
_{\pm}(x)\psi_\lambda$, $x\in \U({\bf sl_2})$.  Then we can define
the Levy process $j_{st}$ over $ \U({\bf sl_2})\otimes \U({\bf
sl_2})$ associated with $(\pi,\eta_\lambda,\tilde \phi_\lambda)$ and
two Levy processes $j_{st}^{\pm}$ over $\U({\bf sl_2})$ associated
with the Schurmann triples $(\pi _+,\eta _{\lambda,+},\tilde \phi
_{\lambda,+})$ and $(\pi_-,\eta _{\lambda,-},\tilde
\phi_{\lambda,-})$ correspondingly such that
$$j_{st}(x)=\Lambda_{st}(\pi (x))+A^*_{st}(\eta_\lambda(x))+A_{st}
(\eta_\lambda(x^*))+(t-s)\phi_\lambda(x)Id,$$
$$ x\in \U({\bf sl_2})\otimes \U({\bf sl_2}),$$
$$j_{st}^{+}(B^{\pm})=\Lambda
_{st}(B_1^{\pm})+A^*_{st}(B^{\pm}_1\psi_\lambda)
+A_{st}(B^{\mp}_1\psi_\lambda)+(t-s)(\psi_\lambda,B^{\pm}_1\psi_\lambda)Id,
$$
$$j_{st}^{+}(M)= \Lambda_{st}(M_1)+A^*_{st}(M_1\psi_\lambda)+A_{st}(M_1
\psi_\lambda)+(t-s)(\psi_\lambda,M_1\psi_\lambda)Id,$$
$$j_{st}^{-}(B^{\pm})=\Lambda _{st}(B^{\pm}_2)+
A^*_{st}(B^{\pm}_2\psi_\lambda)+A_{st}(B^{\mp}_2\psi_\lambda)+
(t-s)(\psi_\lambda,B^{\pm}_2\psi_\lambda)Id,$$
$$j_{st}^{-}(M)= \Lambda _{st}(M_2)+A^*_{st}(M_2\psi_\lambda)+A_{st}(M_2
\psi_\lambda )+ (t-s)(\psi_\lambda,M_2\psi_\lambda)Id.$$
One can associate with the Levy processes given above
two representations $\theta_{\pm}=\theta^{(\lambda)}_\pm$ of the SWN algebra
$\A$
and a representation $\theta $ of the tensor product of
two SWN algebras $\A\otimes \A$ in the same
Fock space $\H=\Gamma (\H_{\phi }\otimes L^2)$ such that
$$\theta _{\pm}(b_{[s,t[})=j_{st}^{\pm}(B^-),\ \theta _{\pm}(b^+_{[s,t[})=
j_{st}^{\pm}(B^+),
$$
$$
\theta _{\pm}(n_{[s,t[})=j_{st}^{\pm}(M)-\gamma (t-s)Id,
$$
$$
\theta (x\otimes y)=\theta _+(x)\theta _-(y),\ x,y\in \A.
$$

{\bf Proposition 4.} {\it The maps $\theta _{\pm}
$ define two representations of the SWN algebra
$\A$ in the Hilbert space
$\H=\Gamma (\H_{\phi }\otimes L^2)$ such that $\theta _+(x)$
and $\theta _-(y)$ are commuting for all $x,y\in \A$.}

Proof.

Notice that $\theta _+(x)$ and $\theta _-(y)$ commute if
$x,y\in \A$ because $B_1^-,B_1^+,M_1$
and $B_2^-,B_2^+,M_2$ commute. Hence we need
to prove only that $\theta _+$ and $\theta _-$ are representations.
Using the CCR relations (see the Introduction) we get
$$[\theta _+(b_{[s,t[}),\theta _+(b^+_{[s,t[})]=\Lambda _{st}
(M_1)+A^*_{st}(M_1\psi_\lambda)+A_{st}(M_1\psi_\lambda)+(\psi_\lambda,M_1
\psi_\lambda)=$$
$$\theta _+(n_{[st[})+\gamma (t-s)Id,$$
$$[\theta _+(n_{[s,t[}),\theta _+(b^{\pm }_{[s,t[})]=\pm \Lambda _{st}
(B^{\pm }_1)\pm A^*_{st}(B_1^{\pm }\psi_\lambda)\mp
A_{st}(B_1^{\mp}\psi_\lambda)\pm(\psi_\lambda,B_1^{\pm }\psi_\lambda)=$$
$$\theta _+(b^{\pm }_{[s,t[}).$$
The remaining formulae can be proved analogously. $\Box$

Define a linear map $\omega _{\lambda }:\A\to \C$,
by the formula
$$
\omega _{\lambda }(x)=(\Omega ,\theta _+(x) \Omega ).
$$

{\bf Theorem 2.} {\it $\omega _{\lambda }$
is a KMS state associated with the evolution $\tau '$,
i.e.
$$
\omega _{\lambda }(xy)=\omega _{\lambda }(x\tau '_i(y))
$$
for all elements $x,y \in \A$ which are analytic with respect
to $\tau '$.}

Proof.

Notice that $\tau '_i(b^{\pm }_{\chi _{[s,t[}})=
\lambda ^{\pm 1}b^{\pm }_{\chi _{[s,t[}}$. Hence
it is sufficient to prove only that
$$
\omega _{\lambda }(b^{\pm}_{\chi _{[s,t[}}y)=
\lambda ^{\pm 1}\omega (yb^{\pm }_{\chi _{[s,t[}}),\ y\in \A.
$$
Given $x=\theta_+(y)\in \theta _+(\A)$ one can obtain
$$
\omega _{\lambda }(b^+_{\chi _{[s,t[}}y)=
(\Lambda _{st}(B^-_1)\Omega ,x\Omega)+$$
$$(A^*_{st}(B^-_1\psi_\lambda)\Omega ,x\Omega )
+(A_{st}(B^+_1\psi_\lambda)\Omega,
x\Omega )+(t-s)(\psi_\lambda,B^+_1\psi_\lambda)\omega _{\lambda }(x)\equiv
{\cal S}.$$
Notice that
$$
(\psi _{\lambda },B_2^+\psi _{\lambda })=(\psi _{\lambda },
{\bf 1}\otimes \rho ^-(B^+)\psi _{\lambda })=
$$
$$
(1-\lambda )(\sum \limits _{n=0}^{+\infty }\lambda ^{\frac {n}{2}}
e_n\otimes e_n,\sum \limits _{n=0}^{+\infty }\lambda ^{\frac {n}{2}}
e_n\otimes \sqrt {n(n+1)}e_{n-1})=0.
$$
Hence
$$
0=(\psi _{\lambda }
,B^+_2\psi _{\lambda })=\overline{(\psi
_{\lambda },B^+_2\psi _{\lambda })} =(
B^+_2\psi _{\lambda },\psi
_{\lambda })=(\psi _{\lambda },B^-_2\psi
_{\lambda }).
$$
Using the last relation, Proposition 4 and the equality
$\Lambda _{st}(x)\Omega =
A_{st}(\xi )\Omega =0,\ x\in \U({\bf sl_2}),\ \xi \in \H$,
we get the following expression,
$$
{\cal S}=\sqrt {\lambda
}((A^*_{st}(B^-_2\psi_\lambda)\Omega , x\Omega
)+(t-s)(\psi_\lambda,B^+_2\psi_\lambda)\omega _{\lambda }(x))=$$
$$=\sqrt {\lambda }((
\Lambda _{st}(B^-_2)\Omega ,x\Omega)+$$
$$(A^*_{st}(B^-_2\psi )\Omega ,x\Omega )+(A_{st}(B^+_2\psi )\Omega ,
x\Omega )+(t-s)(\psi ,B^-_2\psi)\omega _{\lambda }(x))=$$
$$
\sqrt {\lambda }(\theta _-(b_{\chi _{[st[}})\Omega,x\Omega )=
\sqrt {\lambda }(\Omega ,x\theta_-(b^+_{\chi _{[st[}})\Omega )=
\lambda \omega_{\lambda }(yb^+_{\chi _{[s,t[}})\label{kms}
$$
The equality
$$
\omega _{\lambda}(b_{\chi _{[s,t[}}x)=
\frac {1}{\lambda }\omega _{\lambda
}(xb_{\chi _{[s,t[}}),\ x\in \A,
$$
can be checked in the same way.
$\Box$

In every representation $\theta $ of the algebra $\A$ there are
sufficiently many hermitian operators $x\in \theta (\A)$. One can
apply the functions $f\in L^{\infty }$ to these operators and
consider the von Neumann algebra $\M=\theta (\A)''\wedge \B(\H)$
generated by the all $f(x)\in \B(\H)$.
The von Neumann algebra $\M$ generated by
the irreducible representation $\theta $ is
of type $\rm I$. Let $\M=\M_{\pm}$
be the von Neumann algebras
generated by the representations constructed from the KMS state
associated with the evolution $\tau '_{\lambda }$ on $\A$.
These representations are not irreducible. In particular,
the algebras $M_+$ and $M_-$ commute.
One might expect $\M$ to be of type $III_{\lambda },\ 0<\lambda <1$
(see [C73]).\bigskip

\medskip

{\bf Acknowledgments}

Grigori Amosov is grateful to Professor Luigi Accardi for
kind hospitality during
his visit at the Centro Vito Volterra of Universita di Roma Tor Vergata.

\medskip

{\bf References}

[ALV99] L. Accardi, Y.G. Lu, I.V. Volovich, {\it White noise approach
to classical and quantum stochastic calculi}, Centro Vito Volterra,
Universita di Roma ``Tor Vergata'', Preprint 375, 1999.

[AFS00] L. Accardi, U. Franz, M. Skeide, {\it Renormalixzed squares of
white noise and other non-gaussian noises as Levy processes on real Lie
algebras}, Centro Vito Volterra, Universita di Roma ``Tor Vergata'',
Preprint 423, 2000, Commun. Math. Phys. 228 (2002) 123-150.

[Par92] K.R. Parthasarathy, {\it An Introduction to Quantum
Stochastic Calculus}, Birkhauser, 1992.

[Mey95] P.-A. Meyer, {\it Quantum Probability for Probabilitists,
Lecture Notes in Math.}, Vol. 1538, Springer-Verlag, Berlin,
2nd edition, 1995.

[C73] A. Connes, {\it Une classification des facteurs de type III}, Ann.
Sci.
Ecole Norm. Sup. 6 (1973) 133-252.

\end {document}